\documentclass[prb,aps,twocolumn,showpacs,nofootinbib]{revtex4}
\usepackage{epsfig}
\usepackage{hyperref}
\usepackage{graphics}
\usepackage{color}

\begin{document}

\author{Hai-Xiao Xiao$^{1,2}$}~\email[]{Email: xiaohaixiao@126.com}
\author{Jing-Rong Wang$^{3}$}~\email[]{Email: wangjr@hmfl.ac.cn}
\author{Hong-Tao Feng$^{5}$, Pei-Lin Yin$^{5}$ and Hong-Shi Zong$^{1,2,6,7}$}~\email[]{Email: zonghs@nju.edu.cn}

\address{$^{1}$Key Laboratory of Modern Acoustics, MOE, Institute of Acoustics, and Department of Physics, Collaborative Innovation Center of Advanced Microstructures, Nanjing University, Nanjing 210093, China}
\address{$^{2}$Department of Physics, Nanjing University, Nanjing 210093, China}
\address{$^{3}$Anhui Province Key Laboratory of Condensed Matter Physics at Extreme Conditions, High Magnetic Field Laboratory of the Chinese Academy of Science,Hefei 230031, Anhui, China}
\address{$^{5}$Department of Physics, Southeast University, Nanjing 211189, China}
\address{$^{6}$Joint Center for Particle, Nuclear Physics and Cosmology, Nanjing 210093, China}
\address{$^{7}$State Key Laboratory of Theoretical Physics, Institute of Theoretical Physics, CAS, Beijing, 100190, China}

\title{Dynamical gap generation in 2D Dirac semimetal with deformed Dirac cone}

\begin{abstract}
According to the extensive theoretical and experimental
investigations, it is widely accepted that the long-range Coulomb
interaction is too weak to generate a dynamical excitonic gap in
graphene with a perfect Dirac cone. We study the impact of the
deformation of Dirac cone on dynamical gap generation. When a
uniaxial strain is applied to graphene, the Dirac cone is made
elliptical in the equal-energy plane and the fermion velocity
becomes anisotropic. The applied uniaxial strain has two effects: it
decreases the fermion velocity; it increases the velocity
anisotropy. After solving the Dyson-Schwinger gap equation, we show
that dynamical gap generation is promoted by the former effect, but
is suppressed by the latter one. For suspended graphene, we find
that the systems undergoes an excitonic insulating transition when
the strain is roughly 7.34$\%$. We also solve the gap equation in
case the Dirac cone is tiled, which might be realized in the organic
material $\alpha$-(BEDT-TTF)$_{2}$I$_{3}$, and find that the tilt of
Dirac cone can suppress dynamical gap generation. It turns out that
the geometry of the Dirac cone plays an important role in the
formation of excitonic pairing.
\end{abstract}


\maketitle

\section{INTRODUCTION}

Semimetals, no matter topologically trivial or non-trivial, have
attracted intensive theoretical and experimental studies because of
their intriguing properties and promising industrial applications
\cite{Vafek14, Wehling14, Weng16, Yan17, Hasan17}. Among all known
semimetals, two-dimensional (2D) Dirac semimetal plays a special
role. There are two famous examples for such semimetals: graphene
\cite{CastroNeto09, Kotov12}; surface state of three-dimensional
(3D) topological insulator (TI) \cite{Hasan10, Qi11}. The low-energy
excitations of these systems are massless Dirac fermions, described
by the relativistic Dirac equation.

In contrast to normal metals that possess a finite Fermi surface,
the Fermi surface of 2D Dirac semimetal shrinks to a number of
discrete points at which the valence and conduction bands touch
\cite{CastroNeto09, Kotov12}. The Coulomb interaction between Dirac
fermions remains long-ranged since the density of states (DOS)
vanishes at the Fermi level. The influences of long-range Coulomb
interaction on the low-energy dynamics of Dirac fermions has been
extensively investigated \cite{Kotov12}. Although weak Coulomb
interaction is found by renormalization group (RG) analysis to be
marginally irrelevant \cite{Shankar94, Gonzalez94, Gonzalez99,
Son07, Hofmann14, Bauer15, Sharma16}, it gives rise to singular
renormalization of fermion velocity \cite{Gonzalez94, Gonzalez99,
Son07, Hofmann14, Bauer15, Sharma16} and logarithmic-like
corrections to a variety of observable quantities, including
specific heat, optical conductivity, thermal conductivity, and
compressibility \emph{etc} \cite{Kotov12}. The singular
renormalization of fermion velocity has been experimentally
confirmed in suspended graphene \cite{Elias11}, quasi-freestanding
graphene on silicon carbide (SiC) \cite{Siegel11}, and graphene on
boron nitride substrate \cite{Yu13}.

The Coulomb interaction can also induce important non-perturbative
effects. Of particular interest is the possibility of
semimetal-insulator quantum phase transition driven by the
dynamical generation of a finite excitonic gap \cite{Khveshchenko01,
CastroNetoPhysics09, Gorbar02, Khveshchenko04, Liu09,
Khveshchenko09, Gamayun10, Sabio10, Zhang11, Liu11, WangLiu11A,
WangLiu11B, WangLiu12, Popovici13, WangLiu14, Gonzalez15,
Carrington16, Gamayun09, WangJianhui11, Katanin16, Vafek08,
Gonzalez10, Gonzalez12, Drut09A, Drut09B, Drut09C, Armour10,
Armour11, Buividovich12, Ulybyshev13, Smith14, Juan12, Kotikov16}.
Once an excitonic gap is opened, the chiral symmetry, corresponding
to symmetry of sublattices, is dynamically broken \cite{Kotov12,
Khveshchenko01}. The research interest in dynamical gap generation
is twofold. Firstly, acquiring a finite gap broadens the possible
applications of graphene in the design and manufacture of electronic
devices \cite{CastroNetoPhysics09}. Secondly, it is the
condensed-matter counterpart of the concept of dynamical chiral
symmetry breaking \cite{Nambu61, Miransky}.

Several years before monolayer graphene was isolated in laboratory,
Khveshchenko \cite{Khveshchenko01} discussed the possibility of
dynamical gap generation for massless Dirac fermions in 2D,
motivated by the theoretical progress of dynamical chiral symmetry
breaking in (2+1)-dimensional QED. His main result
\cite{Khveshchenko01} is that a finite excitonic gap can be
generated if the Coulomb interaction strength $\alpha$, defined by
\begin{eqnarray}
\alpha = \frac{e^{2}}{v\kappa},
\end{eqnarray}
where $e$ is the electric charge, $v$ fermion velocity, and $\kappa$
dielectric constant, is larger than some threshold $\alpha_c$ for a
fixed fermion flavor $N_{f}$. This interesting result has stimulated
extensive theoretic and numerical studies aimed at finding the
precise value of $\alpha_c$. Calculations performed by
 Dyson-Schwinger (DS) equation \cite{Khveshchenko04, Liu09,
Khveshchenko09, Gamayun10, Zhang11, WangLiu11A}, Bethe-Slapeter (BS)
equation \cite{Gamayun09, WangJianhui11}, RG approach \cite{Vafek08,
Gonzalez12}, and Monte Carlo simulation \cite{Drut09A, Drut09B}
found that the critical value $\alpha_{c}$ falls into the range
$0.79 < \alpha_{c} < 2.16$. For suspended graphene, the interaction
strength is $\alpha \approx 2.16$, whereas for graphene placed on
SiO$_{2}$ substrate, it becomes $\alpha \approx 0.79$
\cite{CastroNetoPhysics09}. It thus indicates that suspended
graphene is an excitonic insulator at zero $T$, but graphene placed
on SiO$_{2}$ remains a semimetal. However, experiments did not find
any evidence for the existence of excitonic insulating phase in
suspended graphene even at very low temperatures \cite{Elias11,
Mayorov12}. In Ref.\cite{WangLiu12}, the authors studied the DS
equation for excitonic gap by incorporating the wave function
renormalization, fermion velocity renormalization, and dynamical gap
generation in an unbiased way, and found that $\alpha_{c} \sim 3.2$
\cite{WangLiu12}. According to this result, the Coulomb interaction
in suspended graphene is too weak to drive the semimetal-insulator
phase transition. This conclusion is well consistent with
experiments \cite{Mayorov12}, and is also confirmed by subsequent
more refined DS equation studies \cite{Gonzalez15, Carrington16}. In
addition, recent Monte Carlo simulations \cite{Ulybyshev13, Smith14}
claimed that, although the Coulomb interaction in suspended graphene
is not strong enough to open an excitonic gap, the $\alpha_{c}$ is
close to $2.16$.

Although careful experiments and elaborate theoretical studies have
already provided strong evidences for the absence of excitonic gap in
intrinsic graphene, there have been several proposals attempting
to realize excitonic insulator in similar semimetals. For example,
it was argued that the Coulomb interaction in an organic material
$\alpha$-(BEDT-TTF)$_{2}$I$_{3}$ might be much stronger than
graphene, because its fermion velocity is about one-tenth of the one
observed in graphene \cite{Monteverde13}. In addition, Triola
\emph{et al.} proposed that the fermion excitations of surface state
of some topological Kondo insulators may have extraordinary small
fermion velocities, which would drive the Coulomb interaction to
fall into the strong coupling regime \cite{Triola15}. Moreover, it
seems viable to strengthen the Coulomb interaction and as such
promote excitonic insulating transition by exerting certain
extrinsic influences. Through Monte Carlo simulations, Tang \emph{et
al.} argued that applying a uniform and isotropic strain by about
15$\%$ can make the Coulomb interaction strong enough to open an
excitonic gap \cite{Tang15}.

Recently, the influence of a uniaxial strain on the properties of
Dirac fermions has been studied. First principle calculations
\cite{Pereira09, Choi10} suggested that the uniaxial strain would
cause the carbon-carbon bond become longer along the direction of
the applied strain and get shorter along its orthogonal direction.
As a consequence, the originally perfect Dirac cone is deformed, and
the fermion velocity along the direction of applied uniaxial strain
decreases, whereas the other component of fermion velocity is made
larger. Sharma \emph{et al.} \cite{Sharma17} investigated the
possibility of dynamical gap generation in graphene with an
anisotropic dispersion, and argued that it is promoted by the
velocity anisotropy induced by the uniaxial strain.

In this paper, we study the influence of uniaxial strain on the
formation of excitonic pairing. It is important to emphasize here
that applying a uniaxial strain to graphene has two effects: first,
it lowers the mean value of fermion velocity; second, it increases the velocity
anisotropy. They might have different effects on dynamical gap
generation. We will address this issue by solving the
self-consistent DS equation for the excitonic gap. We will show
that, the two effects of uniaxial strain are actually competitive
since they have opposite influences on dynamical gap generation.
After carrying out numerical calculations based on three widely
adopted approximations, we obtain the dependence of excitonic gap on
two parameters, namely the effective Coulomb interaction strength
$\alpha = e^{2}/\bar{v}\epsilon$, where $\bar{v} =
\sqrt{v_{x}v_{y}}$, and the velocity ratio $\eta = v_{x}/v_{y}$. We
will show that the dynamical gap generation is enhanced if the mean
value of fermion velocity $\bar{v}$ is lowered, but can be strongly
suppressed when the velocity anisotropy grows. Our conclusion is
qualitatively consistent with Ref.~\cite{WangLiu14}. We will present
a comparison between our results and that reported in
Ref.~\cite{Sharma17}.

Apart from strain-induced anisotropy \cite{Pereira09, Choi10}, the
fermion velocity anisotropy can also be induced by introducing
certain periodic potentials \cite{Park08A, Park08B, Rusponi10}.
Moreover, it is found that the surface state of some topological
insulators, including $\beta$-Ag$_{2}$Te \cite{ZhangWei11} and
$\beta$-HgS \cite{Virot11}, is 2D Dirac semimetal with two unequal
components of fermion velocity. Our results of the impact of
velocity anisotropy on dynamical gap generation are applicable to
these systems.

To gain more quantitative knowledge of the effects caused by
uniaxial strain, we extract an approximated expression for the
fermion velocity from recent first-principle calculations of
uniaxially strained graphene \cite{Choi10}. For suspended graphene,
we find that as the applied uniaxial strain grows the system
undergoes a semimetal-insulator phase transition when the strain
becomes larger than 7.34$\%$. The dependence of dynamical gap on the
magnitude of uniaxial strain is obtained from the solution of gap
equation, which shows that the gap is an increasing function of the
applied strain. We thus see that the enhancement of dynamical gap
generation caused by the decreasing mean velocity dominates over the
suppressing effect produced by the increasing velocity anisotropy.

In addition to the uniaxial strain, the Dirac cone may be deformed
in other ways. For instance, the Dirac cone is known to be tilted in
an organic material $\alpha$-(BEDT-TTF)$_{2}$I$_{3}$ \cite{Tajima09,
Isobe12, Nishine10, Sari14, Trescher15, Proskurin15, Hirata17},
which is regarded as a promising candidate to realize the quantum
phase transition from 2D Dirac semimetal to excitonic insulator
\cite{Monteverde13}. We also study the fate of dynamical gap
generation in such systems, and show that it is suppressed when the
Dirac cone is tilted.

The rest sections of the paper are organized as follows. In
Sec.~\ref{Sec:Model}, we give the model action for 2D Dirac fermions
with anisotropic dispersion. In Sec.~\ref{Sec:GapEquation}, we
derive the DS gap equation, and then solve it by employing three
different approximations. The numerical results are presented and
discussed in Sec.~\ref{Sec:NumResults}. In Sec.~\ref{Sec:Comparing},
we compare our results with a recent work. The direct relation
between strain and gap generation is investigated in
Sec.~\ref{Sec:Uniaxialstrain}. The influence of tilted Dirac cone on
dynamical gap generation is studied in Sec.~\ref{Sec:TiltDiracCone}.
We end the paper with a brief summary in Sec.~\ref{Sec:Summary}.

\section{Model and Feynman rules\label{Sec:Model}}

The massless Dirac fermions with an anisotropic dispersion can be
described by the action
\begin{eqnarray}
S &=& \int dtd^{2}\mathbf{r}\bar{\Psi}_{\sigma}(\mathbf{r})(i\gamma_{0}
\partial_{t}-iv_{x}\gamma_{1}\nabla_{x}-i
v_{y}\gamma_{2}\nabla_{y})\Psi_{\sigma}(\mathbf{r})\nonumber
\\
&& -\frac{1}{2}\int dtdt'd^{2}\mathbf{r}d^{2}\mathbf{r}'
\bar{\Psi}_{\sigma1}(\mathbf{r})\gamma_{0}\Psi_{\sigma1}(\mathbf{r})
\nonumber \\
&&\times U_{0}(t-t',|\mathbf{r}-\mathbf{r}'|)
\bar{\Psi}_{\sigma2}(\mathbf{r}')\gamma_{0}\Psi_{\sigma2}(\mathbf{r}').
\label{Eq:Lagrangian}
\end{eqnarray}
In this action, $\Psi_{\sigma}^{T} =
(\Psi_{Ka\sigma},\Psi_{K'a\sigma},\Psi_{Kb\sigma},\Psi_{K'b\sigma})$
is a four-component spinor field, representing the low-energy Dirac
fermion excitations of graphene, where $a$ and $b$ stand for the two
inequivalent sublattices, and $K$ and $K'$ for two valleys. The
fermion flavor $\sigma=1,2,...N_{f}$, corresponding to the spin
components. Although the physical flavors is $N_f = 2$, in the
following analysis we will consider a general large $N_f$ so as to
perform the $1/N_{f}$ expansion. The $\gamma$-matrices are defined
as $\gamma_{0,1,2} = (\tau_{3},i\tau_{2},-i\tau_{1}) \otimes
\tau_{3}$, where $\tau_{1,2,3}$ are the Pauli matrices.

The free fermion propagator
\begin{eqnarray}
G_{0}(i\omega,\textbf{k})=\frac{1}{-i\omega\gamma_{0} +
v_{x}k_{x}\gamma_{1}+v_{y}k_{y}\gamma_{2}}.\label{Eq:FreeFermionPropagator}
\end{eqnarray}
The bare Coulomb interaction between fermions is
\begin{eqnarray}
\emph{U}_{0}(t,\textbf{r}) =
\frac{e^{2}\delta(t)}{\kappa|\mathbf{r}|},\label{eq2}
\end{eqnarray}
where the dielectric constant $\kappa = \epsilon_{r}\epsilon_{0}$.
$\epsilon_{0}$ is the dielectric constant in vacuum, and
$\epsilon_{r}$ is a parameter determined by substrate. After making
a Fourier transformation, we obtain its expression in the momentum
space:
\begin{eqnarray}
\emph{U}_{0}(\textbf{q}) &=& \frac{e^{2}}{\kappa}\int
\frac{d^2\mathbf{x}}{2\pi}\frac{\exp(-i\mathbf{qr})
\delta(t)}{|\mathbf{r}|} \nonumber \\
&=& \frac{2\pi e^{2} \delta(t)}{\kappa|\mathbf{q}|}.\label{eq3}
\end{eqnarray}
The bare Coulomb interaction will always be dynamically screened by
the collective electron-hole pairs, which is encoded by the
polarization function. To the leading order of $1/N_f$ expansion,
the polarization function is given by
\begin{eqnarray}
\Pi(i\Omega,\textbf{q}) &=& -N_{f}\int\frac{d\omega}{2\pi}
\frac{d^2\textbf{k}}{(2\pi)^{2}} \mathrm{Tr}\left[\gamma_{0}
G_{0}(i\omega,\textbf{k})\gamma_{0}\right.\nonumber
\\
&&\left.\times G_{0}(i\omega + i\Omega,\textbf{k} +
\textbf{q})\right]\nonumber \\
&=&\frac{N_{f}}{8v_{x}v_{y}}\frac{v_{x}^{2}q_{x}^{2} +
v_{y}^{2}q_{y}^{2}}{\sqrt{\Omega^{2}+v_{x}^{2}q_{x}^{2} +
v_{y}^{2}q_{y}^{2}}}.\label{eq4}
\end{eqnarray}
After including this polarization, the dressed Coulomb propagator
can be written as
\begin{eqnarray}
D(i\Omega,\mathbf{q}) = \frac{1}{\frac{\kappa|\mathbf{q}|}{2\pi
e^{2}} + \frac{N_{f}}{8v_{x}v_{y}}\frac{v_{x}^{2}q_{x}^{2} +
v_{y}^{2}q_{y}^{2}}{\sqrt{\Omega^2 + v_{x}^{2}q_{x}^{2} +
v_{y}^{2}q_{y}^{2}}}}. \label{Eq:DressedCoulomb}
\end{eqnarray}

\section{Dyson-Schwinger equation}\label{Sec:GapEquation}

In the presence of Coulomb interaction, the dynamics of Dirac
fermions will be significantly affected. Generically, the dressed
fermion propagator has the form
\begin{eqnarray}
G(i\omega,\mathbf{k}) = \frac{1}{-i\omega A_{0}\gamma_{0} +
v_{x}k_{x}A_{1}\gamma_{1} + v_{y}k_{y}A_{2}\gamma_{2} + m},
\label{Eq:FullFermionPropagator}
\end{eqnarray}
where $A_{0,1,2}\equiv A_{0,1,2}(i\omega,\mathbf{k})$ are the
renormalization functions and $m\equiv m(i\omega,\mathbf{k})$ is the
dynamical fermion gap. The renormalized and free propagators are
connected by the DS equation
\begin{eqnarray}
G^{-1}(i\varepsilon,\mathbf{p}) &=&
G_{0}^{-1}(i\varepsilon,\mathbf{p}) + \int\frac{d\omega}{2\pi}
\frac{d^2\mathbf{k}}{(2\pi)^{2}} \gamma_{0}
G(i\omega,\mathbf{k})\nonumber
\\
&&\times\gamma_{0}\Gamma(i\varepsilon,\mathbf{p};i\omega,\mathbf{k})
D(i(\varepsilon-\omega),\mathbf{p}-\mathbf{k}),
\label{Eq:DysonEqOriginal}
\end{eqnarray}
where $\Gamma(i\varepsilon,\mathbf{p};i\omega,\mathbf{k})$ is the
vertex correction. As demonstrated in Refs.~\cite{WangLiu12,
Gonzalez15, Carrington16}, the functions
$A_{0,1,2}(i\omega,\mathbf{k})$ and the vertex
$\Gamma(i\varepsilon,\mathbf{p};i\omega,\mathbf{k})$ play an important
role in the determination of the precise value of $\alpha_c$. The
purpose of the present work is to examine whether dynamical gap
generation is enhanced or suppressed by the velocity anisotropy. The
qualitative impact of anisotropy actually does not rely on the
precise value of $\alpha_c$. For our purpose, we will retain only
the leading order contribution of the $1/N_{f}$ expansion to the DS
equation \cite{Khveshchenko01, Gorbar02, Khveshchenko04, Liu09}, and
set $A_{0} = A_{1} = A_{2} = 1$. Under this approximation, the Ward
identity requires $\Gamma = 1$. Now it is easy to find that the
dynamical fermion gap $m(i\varepsilon,p_{x},p_{y})$ satisfies the
following nonlinear integral equation:
\begin{widetext}
\begin{eqnarray}
m(i\varepsilon,p_{x},p_{y}) &=& \int\frac{d\omega}{2\pi}
\int\frac{dk_{x}}{2\pi} \int\frac{dk_{y}}{2\pi}
\frac{m(i\omega,k_{x},k_{y})}{\omega^2+v_{x}^{2}k_{x}^2 +
v_{y}^{2}k_{y}^{2} + m^{2}(i\omega,k_{x},k_{y})}
\frac{1}{\frac{|\mathbf{q}|}{\frac{2\pi e^{2}}{\kappa}} +
\frac{N_{f}}{8v_{x}v_{y}}\frac{v_{x}^{2}q_{x}^{2} +
v_{y}^{2}q_{y}^{2}}{\sqrt{\Omega^2 + v_{x}^{2}q_{x}^{2} +
v_{y}^{2}q_{y}^{2}}}},\label{Eq:GapEquationGeneral}
\end{eqnarray}
where $$\Omega=\varepsilon-\omega, \qquad q_{x} = p_{x}-k_{x},
\qquad q_{y} = p_{y}-k_{y}.$$ An apparent fact is that the gap
equations is symmetric under the transformation:
$v_{x}\leftrightarrow v_{y}$. In this paper, we define the effective
strength of Coulomb interaction by $\alpha = e^{2}/\bar{v}\kappa$,
where $\bar{v} = \sqrt{v_{x}v_{y}}$, and the velocity anisotropy as
$\eta = \frac{v_{x}}{v_{y}}$, so that these two parameters, the
interaction strength and the velocity anisotropy can be adjusted
separately. Now the two velocities $v_{x}$ and $v_{y}$ can be
re-expressed by $\bar{v}$ and $\eta$ as follows:
\begin{eqnarray}
v_{x} = \sqrt{\eta}\times\bar{v},\qquad v_{y} = \bar{v}/\sqrt{\eta}.
\end{eqnarray}
The above DS gap equation becomes
\begin{eqnarray}
m(i\varepsilon,p_{x},p_{y})&=& \int\frac{d\omega}{2\pi}
\int\frac{dk_{x}}{2\pi} \int\frac{dk_{y}}{2\pi}
\frac{m(i\omega,k_{y},k_{y})}{\omega^2+\eta\bar{v}^{2}k_{x}^2 +
\frac{\bar{v}^{2}k_{y}^{2}}{\eta} + m^{2}(i\omega,k_{x},k_{y})}
\frac{1}{\frac{|\mathbf{q}|}{2\pi\alpha \bar{v}} +
\frac{N_{f}}{8\bar{v}^{2}}\frac{\eta\bar{v}^{2}q_{x}^{2} + 1/\eta
\bar{v}^{2}q_{y}^{2}}{\sqrt{\Omega^2+\eta\bar{v}^{2}q_{x}^{2} +
1/\eta\bar{v}^{2}q_{y}^{2}}}}.\label{Eq:GapEquationGeneralTranform}
\end{eqnarray}
Due to separate dependence of $m$ on the energy and two components
of momenta, it is still very difficult to numerically solve this
nonlinear integral equation. In order to simplify numerical works,
we will employ three frequently used approximations.

\subsection{Hartree-Fock approxiamtion}

Under Hartree-Fock (HF) approximation, the polarization function in
the dressed Coulomb interaction is completely discarded
\cite{WangJianhui11, Sharma17}. Namely, the bare Coulomb interaction
is actually used. Under HF approximation, the gap equation becomes
\begin{eqnarray}
m(p_{x},p_{y})&=&\frac{1}{2} \int\frac{dk_{x}}{2\pi}
\int\frac{dk_{y}}{2\pi}
\frac{m(k_{x},k_{y})}{\sqrt{\eta\bar{v}^{2}k_{x}^2 +
\frac{\bar{v}^{2}k_{y}^{2}}{\eta} +
m^{2}(k_{x},k_{y})}}\frac{1}{\frac{|\mathbf{q}|}{2\pi \alpha
\bar{v}} }.\label{Eq:GapEquationHF}
\end{eqnarray}

\subsection{Instantaneous approximation}

Under instantaneous approximation, the dressed Coulomb interaction
takes the form \cite{Khveshchenko01, Gorbar02}
\begin{eqnarray}
D(i\Omega,\mathbf{q})\rightarrow D(0,\mathbf{q}).
\end{eqnarray}
Accordingly, the gap loses the energy dependence, and depends only
on the momentum. After carrying out the integration over $\omega$,
the gap equation in instantaneous approximation is given by
\begin{eqnarray}
m(p_{x},p_{y}) &=& \frac{1}{2} \int\frac{dk_{x}}{2\pi}
\int\frac{dk_{y}}{2\pi}
\frac{m(k_{x},k_{y})}{\sqrt{\eta\bar{v}^{2}k_{x}^2 +
\frac{\bar{v}^{2}k_{y}^{2}}{\eta} +
m^{2}(k_{x},k_{y})}}\frac{1}{\frac{|\mathbf{q}|}{\frac{2\pi
e^{2}}{\kappa}} + \frac{N_{f}}{8\bar{v}}\sqrt{\eta
q_{x}^{2}+\frac{q_{y}^{2}}{\eta}}}.\label{Eq:GapEquationInstantaneous}
\end{eqnarray}

\subsection{Gamayun-Gorbar-Guysin (GGG) approximation}

It is well known that dynamical screening plays a crucial role in the
determination of the effective strength of Coulomb interaction
\cite{Gamayun10}. In an approximation proposed by Gamayun, Gorbar,
and Gusynin (GGG) \cite{Gamayun10}, the dynamical screening of
Coulomb interaction is partially considered. In the GGG
approximation, the gap $m(i\varepsilon,\textbf{p})$ is supposed to
be energy-independent, i.e.,
\begin{eqnarray}
m(i\varepsilon,p_{x},p_{y})\rightarrow m(p_{x},p_{y}),\nonumber
\end{eqnarray}
but the energy dependence of the polarization is explicitly
retained. Applying this approximation leads to
\begin{eqnarray}
m(p_{x},p_{y}) &=& \int\frac{d\omega}{2\pi} \int\frac{dk_{x}}{2\pi}
\int\frac{dk_{y}}{2\pi}\frac{m(k_{x},k_{y})}{\omega^2+\eta
\bar{v}^{2}k_{x}^2 + \frac{\bar{v}^{2}k_{y}^{2}}{\eta} +
m^{2}(k_{x},k_{y})}\frac{1}{\frac{|\mathbf{q}|}{2\pi\alpha\bar{v}} +
\frac{N_{f}}{8\bar{v}^{2}}\frac{\eta
\bar{v}^{2}q_{x}^{2}+\frac{\bar{v}^{2}q_{y}^{2}}{\eta}}{\sqrt{\omega^2
+ \eta\bar{v}^{2}q_{x}^{2} +
\frac{\bar{v}^{2}q_{y}^{2}}{\eta}}}}.\label{eq12}
\end{eqnarray}
Performing the integration of $\omega$, the gap equation can be
further written as
\begin{eqnarray}
m(p_{x}',p_{y}')&=&\alpha\int\frac{dk_{x}}{2\pi}
\int\frac{dk_{y}'}{2\pi}\frac{m(k_{x}',k_{y}')}{\sqrt{\eta k_{x}'^2
+ \frac{k_{y}'^{2}}{\eta}+m^{2}(k_{x}',k_{y}')}}
\frac{J(d,g)}{\sqrt{\left(p_{x}'-k_{x}'\right)^2+\left(p_{y}'-k_{y}'\right)^2}},
\label{eq13}
\end{eqnarray}
\end{widetext}
where we have employed the transformations
\begin{eqnarray}
&& \bar{v}p_{x}\rightarrow p_{x}', \quad \bar{v}p_{y}\rightarrow
p_{y}',\nonumber \\
&& \bar{v}k_{x}\rightarrow k_{x}', \quad \bar{v}k_{y}\rightarrow
k_{y}'.
\end{eqnarray}
The function $J(d,g)$ is given by
\begin{eqnarray}
J(d,g) = \frac{\left(d^{2} - 1\right)\left[\pi - g c(d)\right] +
dg^{2}c(g)}{d^{2}+g^{2}-1},\label{Eq:JdgDef}
\end{eqnarray}
where
\begin{eqnarray}
c(x)=\left\{
\begin{array}{ll}
\frac{2}{\sqrt{1-x^{2}}}\cos^{-1}\left(x\right) & x<1
\\
\\
\frac{2}{\sqrt{x^{2}-1}}\cosh^{-1}\left(x\right) & x>1
\\
\\
2 & x=1
\end{array}\label{Eq:CxDef}
\right.,
\end{eqnarray}
and
\begin{eqnarray}
d&=&\sqrt{\frac{\eta k_{x}'^2 +
\frac{k_{y}'^{2}}{\eta} +
m^{2}(k_{x}',k_{y}')}{\eta\left(p_{x}'-k_{x}'\right)'^{2}
+\frac{\left(p_{y}'-k_{y}'\right)^{2}}{\eta}}}\label{Eq:dDef}
\\
g&=& \frac{N_{f}\pi \alpha\sqrt{\eta \left(p_{x}' -
k_{x}'\right)^{2} + \frac{\left(p_{y}' -
k_{y}'\right)^{2}}{\eta}}}{4\sqrt{\left(p_{x}'-k_{x}'\right)^2 +
\left(p_{y}'-k_{y}'\right)^2}}.\label{Eq:gDef}
\end{eqnarray}

\begin{figure}
\includegraphics[width=2.5in]{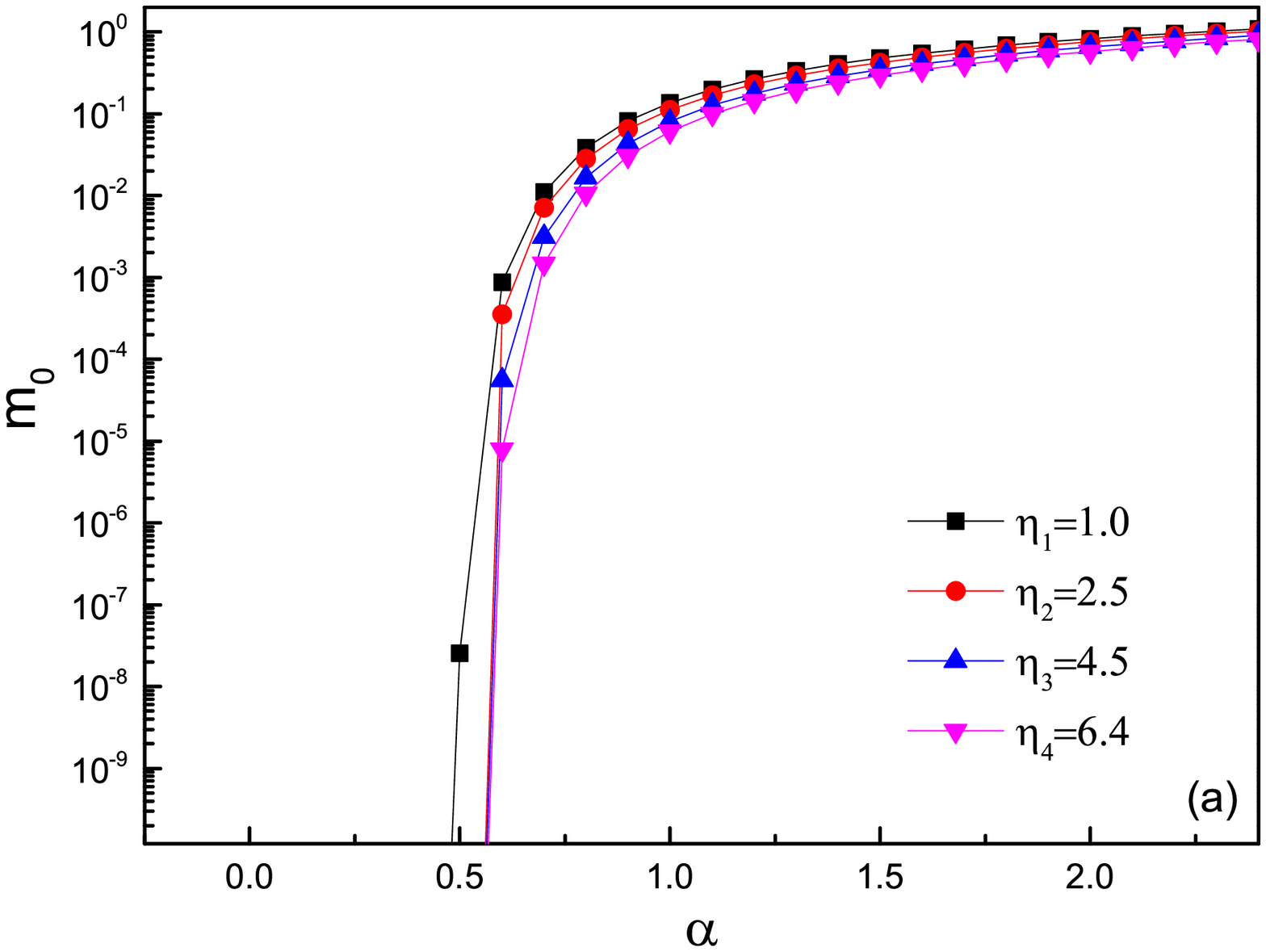}
\includegraphics[width=2.5in]{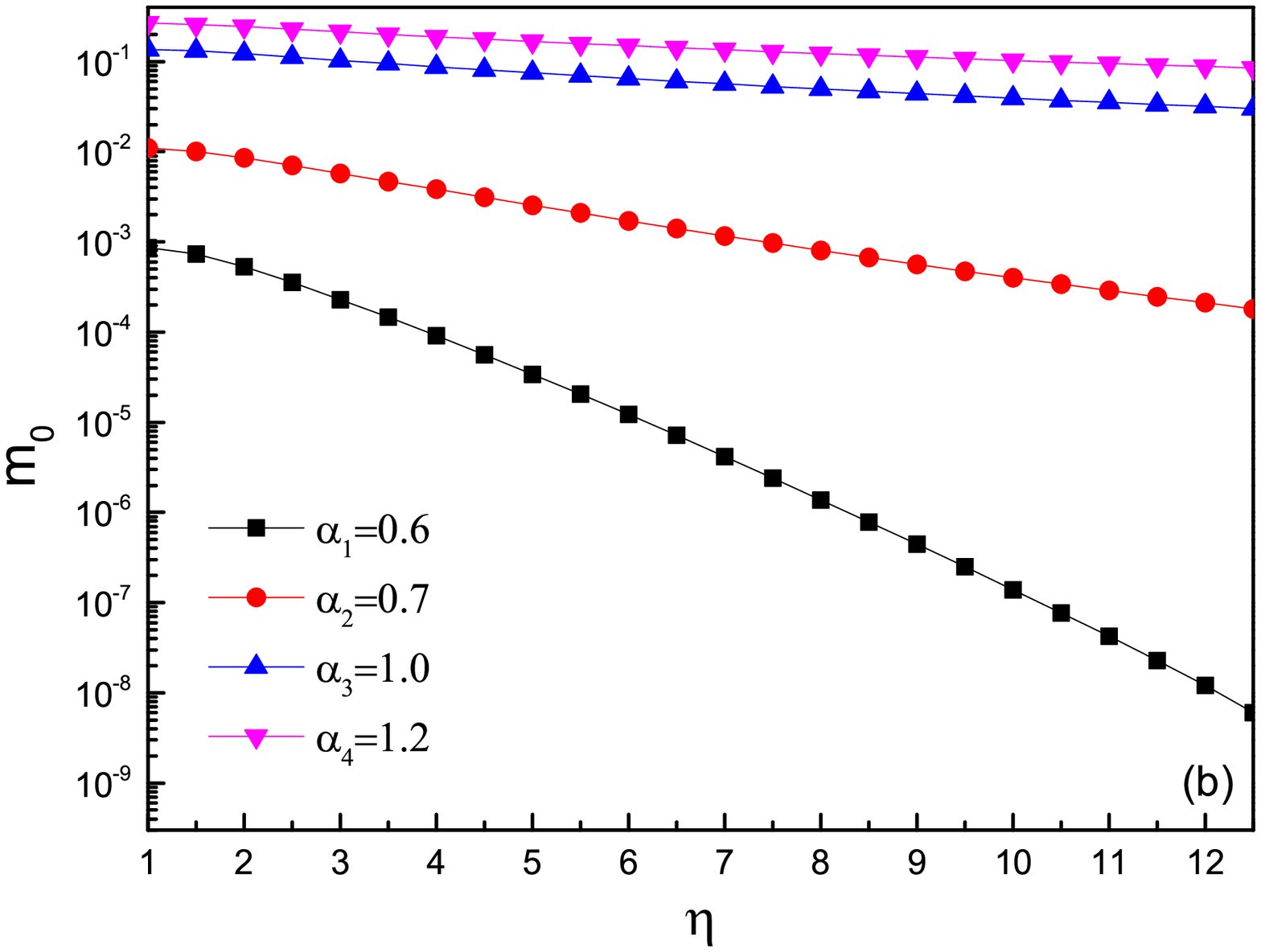}
\caption{(a) $\alpha$ dependence of $m_{0}$ for different $\eta$;
(b) $\eta$ dependence of $m_{0}$ for different $\alpha$. HF
approximation is employed.}\label{Fig:Gap0HartreeFock}
\end{figure}

\begin{figure}
\includegraphics[width=2.5in]{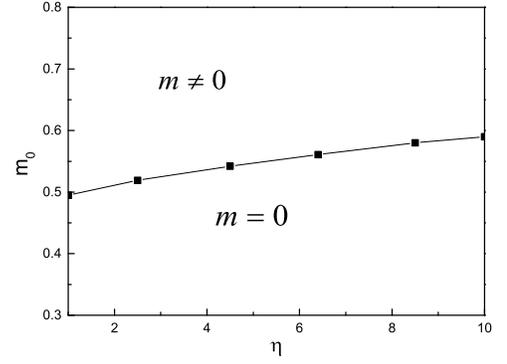}
\caption{Phase diagram on $\alpha$-$\eta$ plane under HF
approximation. \label{Fig:PhaseDiagramHatreeFock}}
\end{figure}

\section{NUMERICAL RESULTS}\label{Sec:NumResults}

We utilize the iteration method to solve the DS gap equation
numerically. To ensure the reliability of numerical results, we have
chosen a series of different initial iteration values. As the gap
equation is symmetry under $\eta \rightarrow \frac{1}{\eta}$, we
only consider the case of $\eta > 1$. The energy scale we adopt here
and following , without special mention, is $\bar{v}\Lambda$, here
$\Lambda$ is the energy cutoff of momentum integration of mass gap
equation.

\subsection{Hartree-Fock approximation}

We first consider the HF approximation. The relation between
$m_{0}\equiv m(0,0)$ and $\alpha$ obtained at a series of different
values of $\eta$ is displayed in Fig.~\ref{Fig:Gap0HartreeFock}(a). In
the isotropic case with $\eta = 1$, we find that the critical value
for dynamical gap generation is roughly $\alpha_c \approx 0.5$,
which is consistent with Ref.~\cite{WangJianhui11}. According to
Fig.~\ref{Fig:Gap0HartreeFock}(b), $m_{0}$ decreases with increasing of
fermion velocity anisotropy. As displayed in
Fig.~\ref{Fig:PhaseDiagramHatreeFock}, in the parameter space of
$\alpha$ and $\eta$, the semimetal phase is enlarged, but the
excitonic insulating phase is compressed with the increasing of
fermion velocity anisotropy. These results show that fermion
velocity anisotropy suppresses dynamical gap generation, which is in
contrast with the conclusion under HF approximation given by
Ref.~\cite{Sharma17}

\subsection{Instantaneous approximation}

\begin{figure}
\includegraphics[width=2.5in]{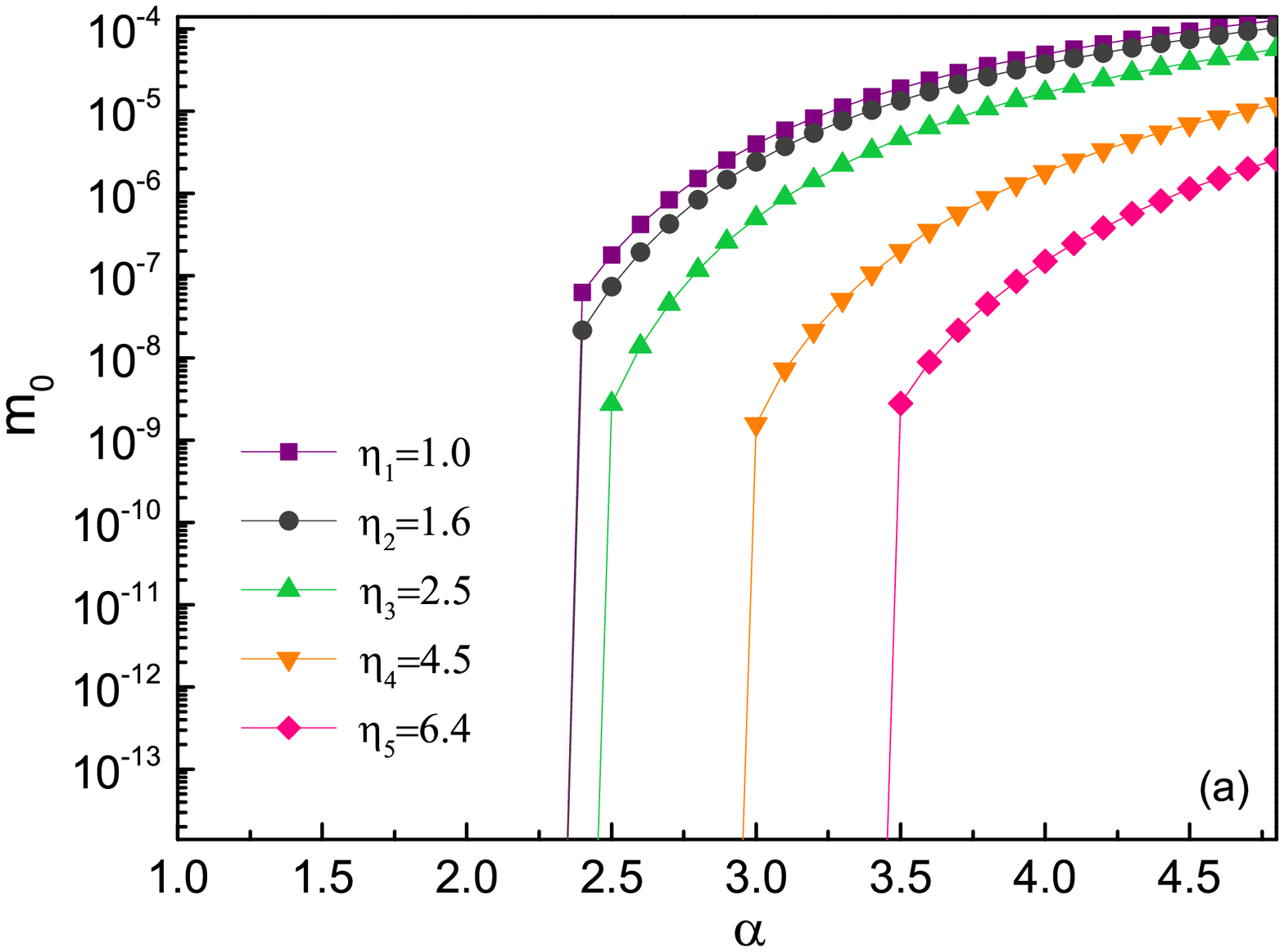}
\includegraphics[width=2.5in]{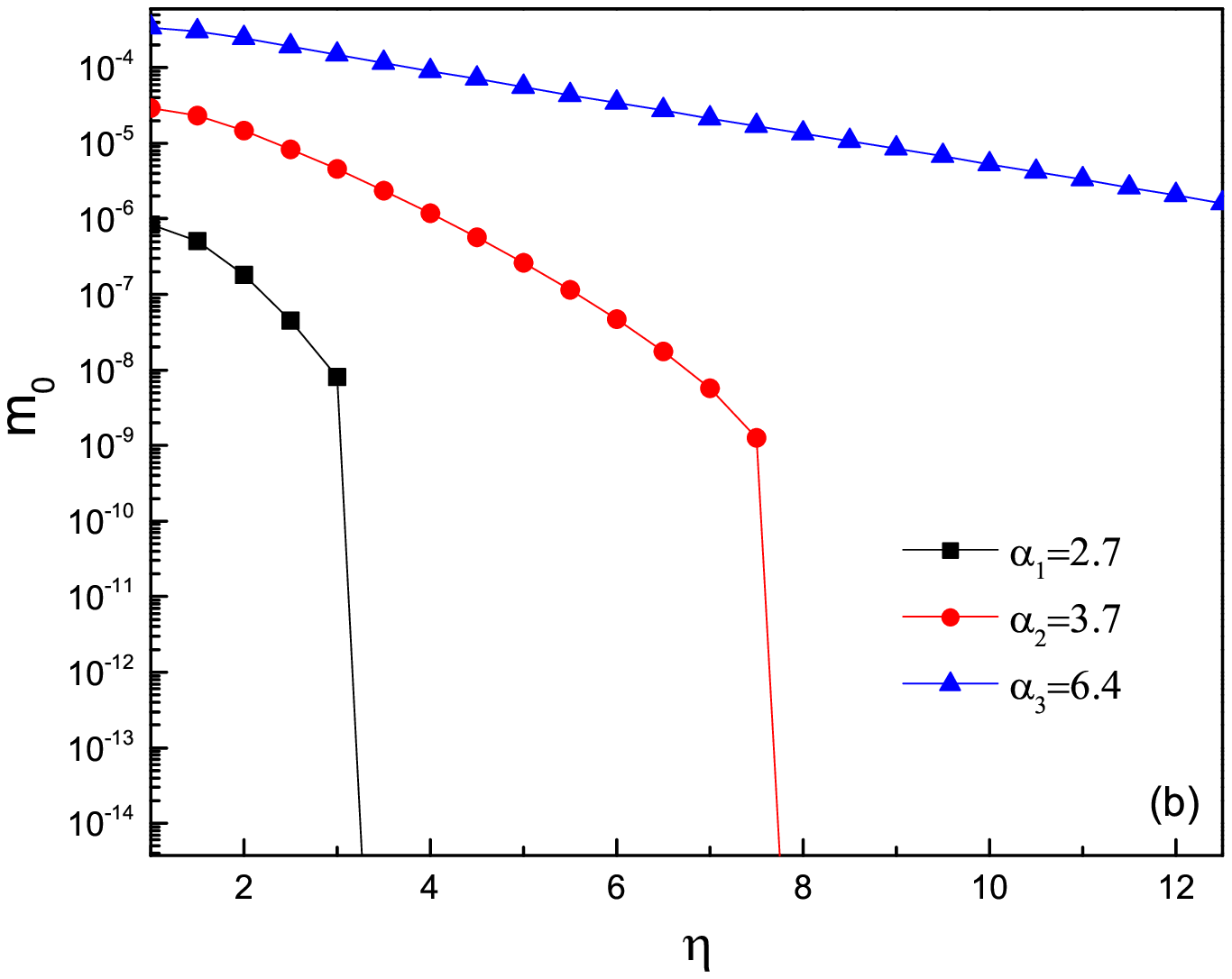}
\caption{(a) Dependence of $m_{0}$ on $\alpha$ with different
$\eta$; (b) Dependence of $m_{0}$ on $\eta$ with different $\alpha$.
Instantaneous approximation is taken. $N_{f}=2$ is taken in this figure, and
Figs.~\ref{Fig:PhaseDiagramInstan}-\ref{Fig:PhaseDiagramTilted}.
\label{Fig:Gap0Instan}}
\end{figure}

The dependence of zero-energy gap $m_{0}$ on $\alpha$ is shown in
Fig.~\ref{Fig:Gap0Instan}(a), where several values of $\eta$ are
assumed. For $\eta = 1$, we find that $\alpha_{c} \approx 2.33$,
which is in good agreement with previous results obtained under the
instantaneous approximation \cite{Khveshchenko01, Gorbar02}. The
critical value $\alpha_{c}$ under instantaneous approximation is
obviously larger than the one under HF approximation, which
indicates that the screening from polarization suppresses the
dynamical gap generation obviously. As can be observed from
Fig.~\ref{Fig:Gap0Instan}(a), $m_{0}$ increases monotonously as
$\alpha$ grows. As $\eta$ becomes larger, $m_{0}$ reduces and
$\alpha_{c}$ increases.

Then we set the value of $\alpha$ to study the anisotropy
dependence of excitonic mass gap. The results are shown in
Fig.~\ref{Fig:Gap0Instan}(b). As we can see, for different values
of interaction strength, the increase of velocity anisotropy
suppresses the formation of dynamical generated gap. Also there
is a critical velocity anisotropy $\eta_{c}$, above which the
excitonic gap dismisses.

\begin{figure}
\includegraphics[width=2.8in]{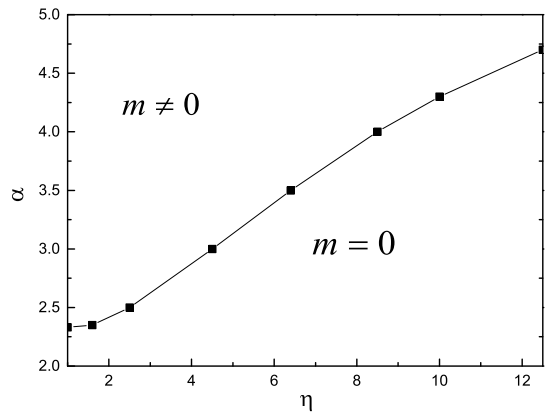}
\caption{Phase diagram on $\alpha$-$\eta$ plane under instantaneous
approximation.  \label{Fig:PhaseDiagramInstan}}
\end{figure}

A schematic phase diagram is depicted on the $\alpha$-$\eta$ plane,
shown in Fig.~\ref{Fig:PhaseDiagramInstan}. There is a critical line
between the semimetal phase with vanishing $m_{0}$ and the excitonic
insulating phase where $m_{0} \neq 0$. This phase diagram tells us that
the dynamical gap can be more easily generated by stronger Coulomb
interaction and smaller velocity anisotropy.

\begin{figure}
\includegraphics[width=2.8in]{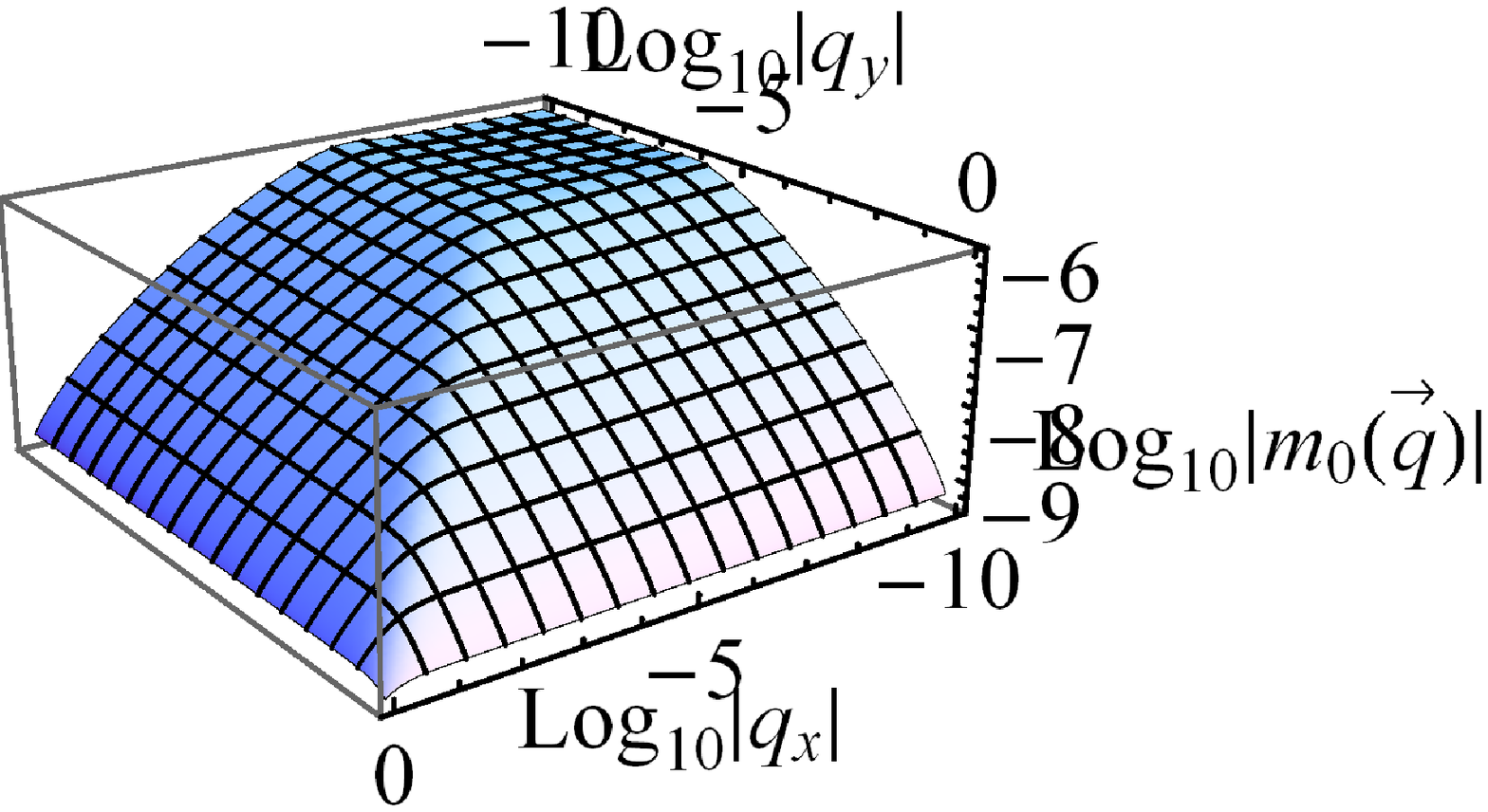}
\includegraphics[width=2.8in]{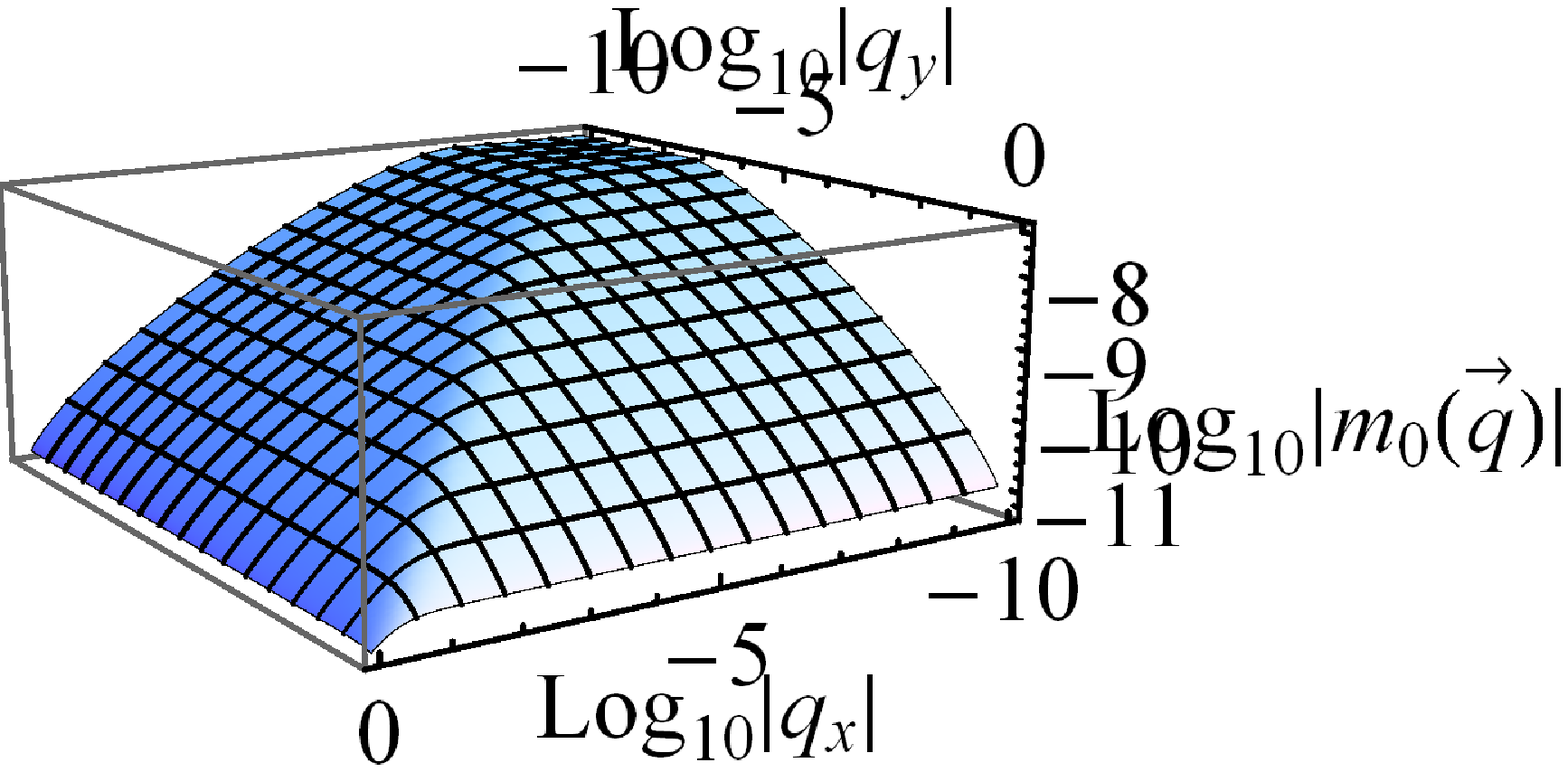}
\caption{Momentum dependence of mass gap for (a) $\eta=1.0$ and (b)
$\eta=2.5$. $\alpha=2.7$ is taken.
\label{Fig:GapMomentumInstan}}
\end{figure}

The dependence of dynamical gap $m(p_{x},p_{y})$ on the momentum
components $|p_{x}|$ and $|p_{y}|$ is displayed in the
Fig.~\ref{Fig:GapMomentumInstan}. Here we choose $\alpha = 2.7$. The
results obtained at $\eta = 1$ and $\eta = 2.5$ are presented in
Figs.~\ref{Fig:GapMomentumInstan} (a) and (b), respectively. In the
isotropic case, $m(p_{x},p_{y})$ is symmetric under the
transformation $p_{x}\leftrightarrow p_{y}$. However, as shown in
Fig.~\ref{Fig:GapMomentumInstan}(b), the anisotropy of fermion
velocities breaks this symmetry.

\subsection{Gamayun-Gorbar-Guysin (GGG) approximation}

\begin{figure}
\includegraphics[width=2.8in]{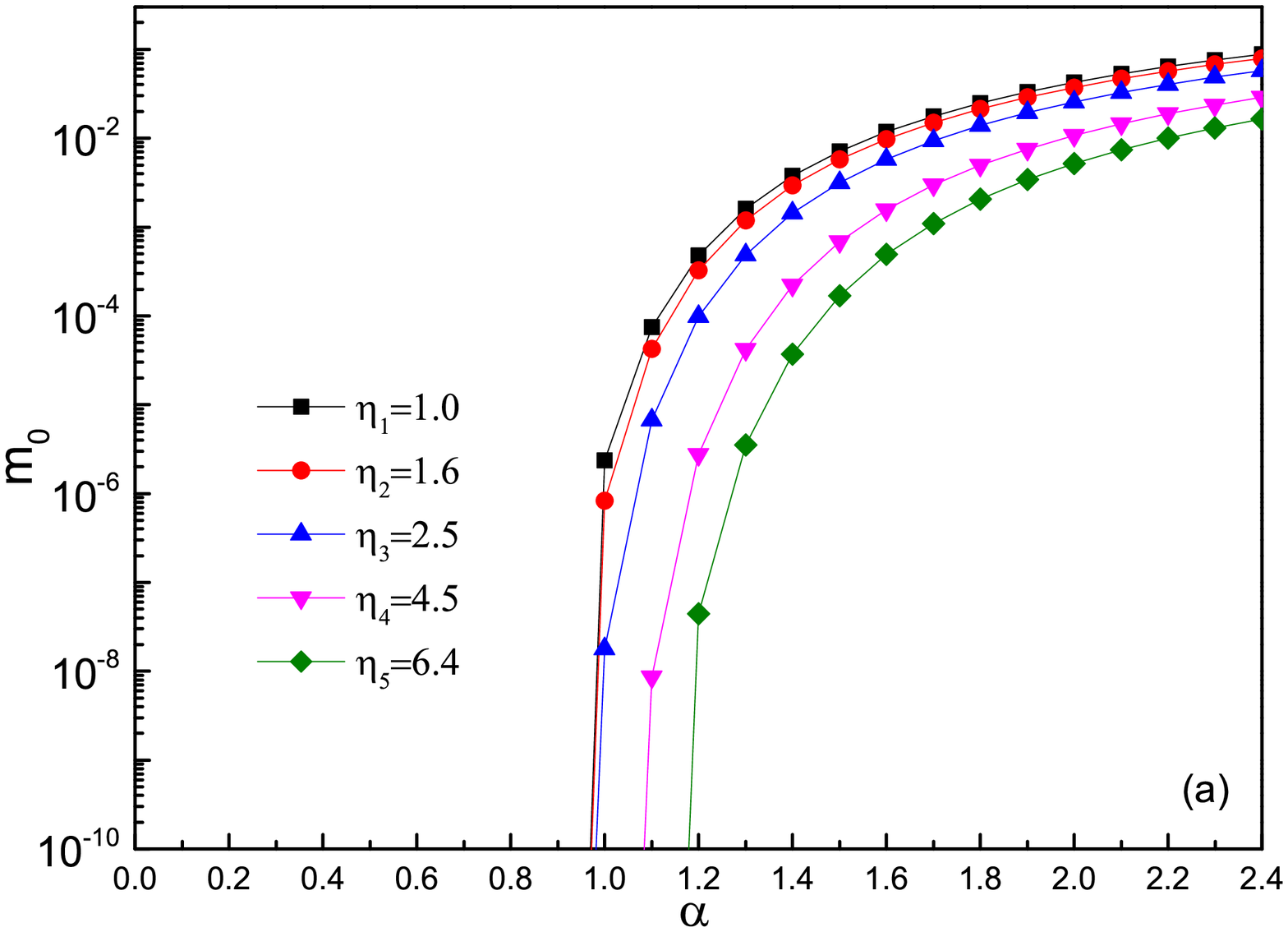}
\includegraphics[width=2.8in]{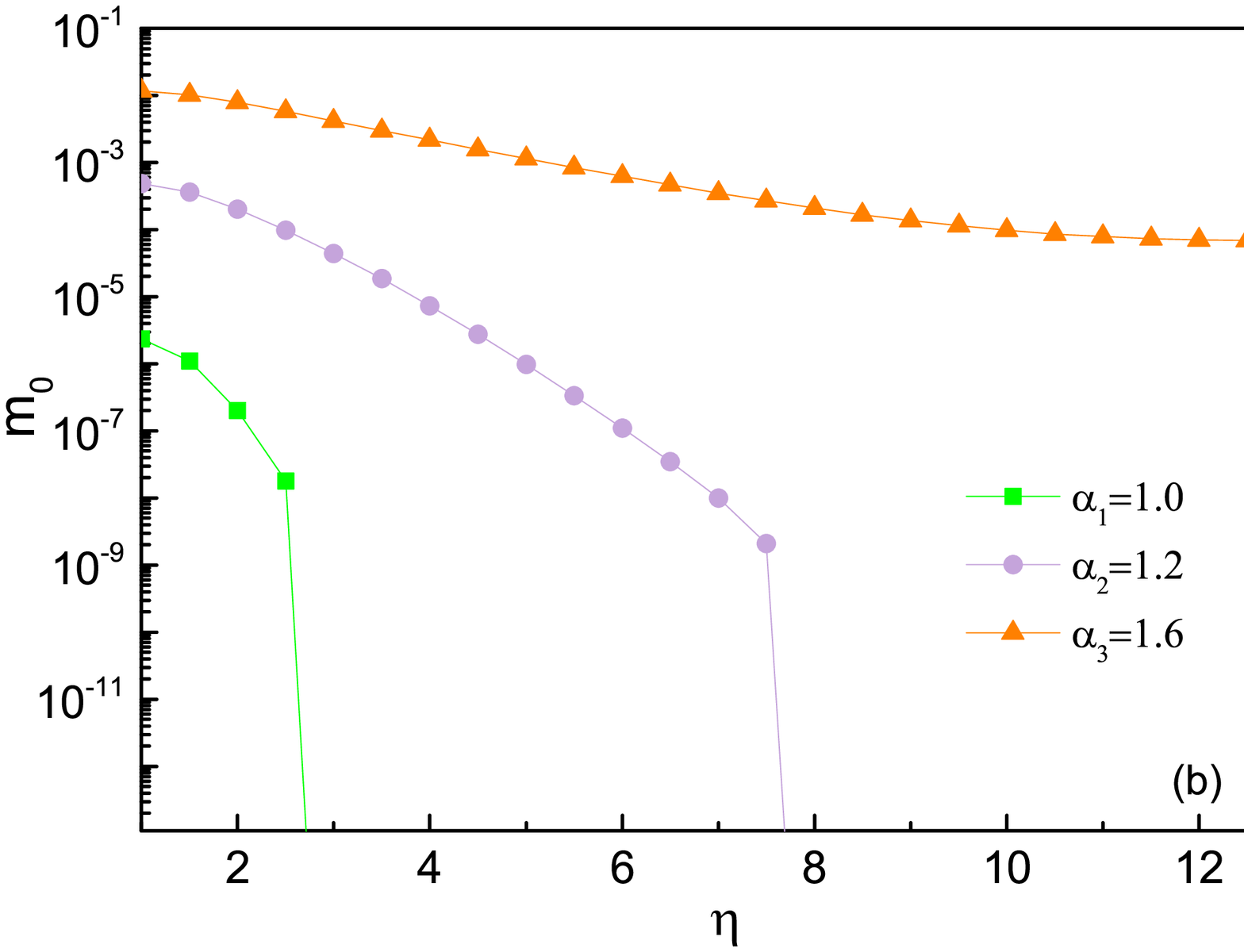}
\caption{(a) $\alpha$ dependence $m_{0}$ for different $\eta$. (b)
$\eta$ dependence of $m_{0}$ for different $\alpha$. GGG
approximation is taken.  \label{Fig:Gap0GGG}}
\end{figure}

For several different values of $\eta$, the curves for the
dependence of $m_{0}$ on the Coulomb strength $\alpha$ within GGG
approximation are depicted in Fig.~\ref{Fig:Gap0GGG}(a). We can see
that there is a critical interaction strength $\alpha_{c}$ above
which a finite excitonic gap can be dynamically generated, the
magnitude of dynamically generated gap increases with the increasing
of interaction strength. It indicates dynamical gap generation
appears only if the Coulomb interaction is strong enough. For the
isotropic case, we get a critical interaction strength
$\alpha_{c}=0.92$, which is in accordance with
Ref.~\cite{Gamayun10}. This critical Coulomb strength is much
smaller than the one obtained within instantaneous approximation
$\alpha_{c}\approx2.33$, which reflects that energy dependence of
dressed Coulomb interaction promotes the dynamical gap generation.
As the velocity anisotropy increases, the magnitude of dynamical gap
decreases, and the critical interaction strength increases, which is
consistent with the results of instantaneous approximation.

Dependence of $m_{0}$ on the fermion velocity anisotropy $\eta$ with
three different values of $\alpha$ is presented in
Fig.~\ref{Fig:Gap0GGG}(b). It is easy to find that $m_{0}$ decreases
monotonously with growing $\eta$, and is completely suppressed when
$\eta$ is larger than a critical value.

\begin{figure}
\includegraphics[width=2.8in]{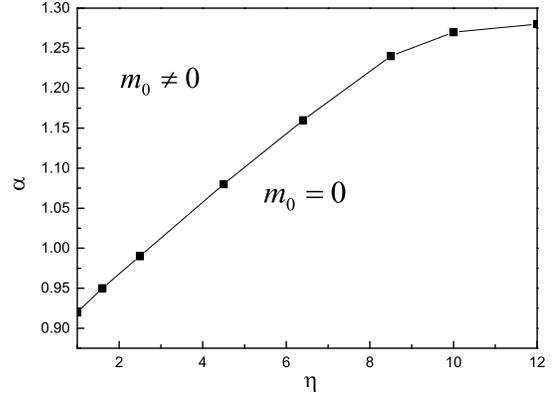}
\caption{Phase diagram of $\alpha$ and $\eta$ under GGG
approximation. }\label{Fig:PhaseDiagramGGG}
\end{figure}

Finally, we give the phase diagram in the $\alpha$-$\eta$ plane in
Fig.~\ref{Fig:PhaseDiagramGGG}. The qualitative characteristic of
Fig.~\ref{Fig:PhaseDiagramGGG} is nearly the same as
Figs.~\ref{Fig:PhaseDiagramHatreeFock} and
\ref{Fig:PhaseDiagramInstan}. All of the three phase diagrams show
that increasing the velocity anisotropy leads to a suppression of
dynamical gap generation.

\section{Comparison with recent work \label{Sec:Comparing}}

In an ideal graphene, the fermions have a universal velocity $v$.
Under certain circumstances, there might be an anisotropy and the
fermion velocity take different values in different directions. For
an isotropic graphene, the velocity anisotropy can be induced by
applying uniaxial strain or other manipulations \cite{Pereira09, Choi10}.

In a recent work, Sharma \emph{et al.} \cite{Sharma17} studied the
dynamical gap generation in a uniaxially strained graphene. They
have introduced two parameters, namely
\begin{eqnarray}
\alpha_{x} = \frac{e^{2}}{\kappa v_{x}}
\end{eqnarray}
and
\begin{eqnarray}
\eta'=\frac{1}{\eta}=\frac{v_{y}}{v_{x}}
\end{eqnarray}
to characterize the interaction strength and velocity anisotropy.
After solving the DS gap equation under the HF and instantaneous
approximations, they concluded that at a fixed $\alpha_{x}$, the
magnitude of excitonic gap increases monotonously as the parameter
$\eta'$ decreases from the isotropic case $\eta' = 1$. They also
claimed that the critical value $\alpha_{x}^{c}$ for dynamical gap
generation decreases with deceasing $\eta'$. Based on these results,
Sharma \emph{et al.} argued that velocity anisotropy is able to
promote dynamical gap generation.

We would point out that the analysis made by Sharma \emph{et al.} is
problematic. When the parameter $\alpha_{x}$ is fixed at certain
value, the velocity component $v_{x}$ is also fixed. In this case,
there are actually two physical effects when lowering $\eta' =
v_{y}/v_{x}$: first, the velocity component $v_{y}$ decreases;
second;, the fermion velocity anisotropy increases. Thus the
conclusion that velocity anisotropy supports the formation of
excitonic mass actually is a combining effect of lowering the
fermion velocity of $v_{y}$ and increasing of fermion velocity anisotropy. In this sense,
such conclusion is misleading. According to our calculations,
dynamical gap generation is promoted when the fermion velocity
decreases, but is suppressed as the anisotropy is enhanced. Therefore,
the correct interpretation of the results obtained by Sharma
\emph{et al.} \cite{Sharma17} should be: for a fixed $\alpha_{x}$,
the promotion of dynamical gap generation caused by decreasing
velocity is more important than the suppression caused by the growth
of velocity anisotropy.

\section{Effects of uniaxial strain on excitonic gap}\label{Sec:Uniaxialstrain}

\begin{figure}
\includegraphics[width=2.8in]{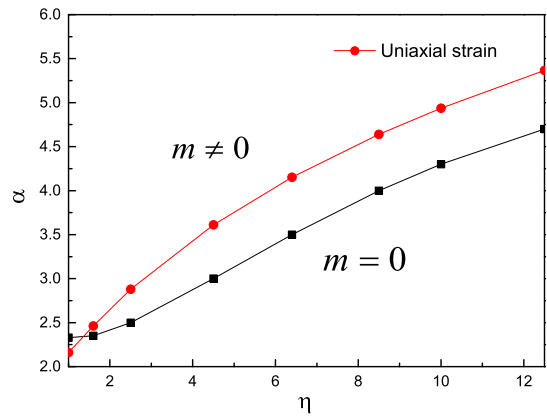}
\caption{Uniaxial strain effects on interaction strength and velocity anisotropy, 
combined with phase diagram in the $\alpha-\eta$ plane, obtained under
the instantaneous approximation. The red lines
represents the trajectory of interaction strength and velocity anisotropy of
suspended graphene under uniaxial strain, and the black line stands for
the critical lines}\label{Fig:Strain-phasediagram}
\end{figure}

The fermion velocities of uniaxially strain graphene can be obtained
by making first-principle calculations \cite{Choi10}. It was found
\cite{Choi10} that the velocities in the $x$- and $y$-directions
vary approximately linearly in strain if the magnitude of strain is
smaller than $24\%$. This result is valid when the strain is applied
in both $A$ and $Z$ directions \cite{Choi10}. For strain
$\varepsilon\%$ in $Z$ direction, we can approximate the fermion
velocities by the following expressions
\begin{eqnarray}
v_{x}&=&v_{0}(1+\frac{1}{120}\varepsilon),
\\
v_{y}&=&v_{0}(1-\frac{7}{240}\varepsilon).
\end{eqnarray}
Accordingly, the velocity anisotropy parameter $\eta$ and the
interaction strength $\alpha$ are re-written as
\begin{eqnarray}
\eta &=& \frac{1+\frac{1}{120}\varepsilon}{1 -
\frac{7}{240}\varepsilon},\label{Eq:Anisotropystrain}
\\
\alpha&=&\frac{e^{2}}{\kappa v_{0}\sqrt{1 -
\frac{5\varepsilon}{240}-\frac{7\varepsilon^{2}}{28800}}}\nonumber
\\
&=&\alpha_{0}\frac{1}{\sqrt{1-\frac{5\varepsilon}{240} -
\frac{7\varepsilon^{2}}{28800}}},\label{Eq:Interactionstrain}
\end{eqnarray}
where $v_{0}$ and $\alpha_{0}$ are the values for graphene without
strain. We suppose $\alpha_{0} \approx 2.2$ for suspended graphene.
It is easy to verify that both $\alpha$ and $\eta$ increase as the
strain grows. Taking advantage of these two relations and the phase diagram in
$\alpha-\eta$ plane, we obtain Fig.~\ref{Fig:Strain-phasediagram}, where the red lines
represents the trajectory of interaction strength and velocity anisotropy of
suspended graphene under uniaxial strain, and the black line stands for
the critical lines on $\alpha-\eta$ plane within instantaneous approximation,
respectively. It is clear that, as the uniaxial strain
increases, an excitonic gap can be dynamically generated.

\begin{figure}
\includegraphics[width=2.8in]{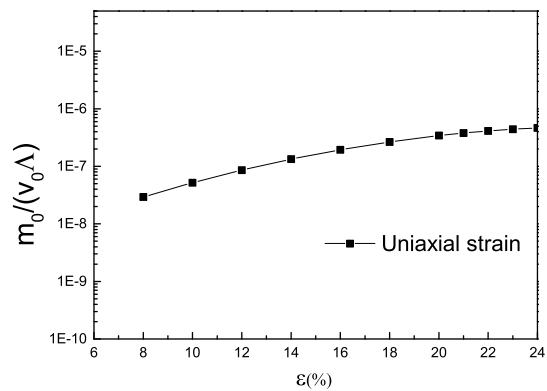}
\caption{Relation between zero-energy gap $m_{0}$ and strain
parameter $\varepsilon$.}\label{Fig:StrainGap}
\end{figure}

The relation between dynamical gap and strain parameter
$\varepsilon$ is presented in Fig.~\ref{Fig:StrainGap}.
The instantaneous approximation is employed in this calculation. In
order to compare $m_{0}$ for different interaction strength, we
adopt a energy scale of the isotropic case, namely $v_{0}\Lambda$ as
the unit of $m_{0}$

The system is gapless when the uniaxial strain is smaller than
7.34$\%$, and a finite dynamical gap is generated as $\varepsilon$
exceeds this critical value. The zero-energy gap $m_{0}$ increases
as $\varepsilon$ grows. This implies that applying a uniaxial strain
to graphene is in favor of excitonic gap generation, and also that
the enhancement effect caused by the decreasing mean velocity
dominates over the suppression effect caused by the increasing
velocity anisotropy. We notice that the magnitude of dynamical gap
$m_{0}$ induced by uniaxial strain is very small, as clearly shown
in Fig.~\ref{Fig:StrainGap}. Therefore, it would be very difficult
to observe such a gap in realistic materials.

\section{2D Dirac semimetal with a tilted cone \label{Sec:TiltDiracCone}}

It was recently argued \cite{Monteverde13} that the organic material
$\alpha$-(BEDT-TTF)$_{2}$I$_{3}$ might be close to a quantum phase
transition between semimetallic and excitonic insulating phases due
to the smallness of the fermion velocity. In the semimetallic phase
of $\alpha$-(BEDT-TTF)$_{2}$I$_{3}$, the Dirac cone is tilted
\cite{Tajima09, Isobe12, Nishine10, Sari14, Trescher15, Proskurin15,
Hirata17}, which can be considered as a deformation of the perfect
Dirac cone realized in an intrinsic graphene. In this section, we
examine whether the tilt of Dirac cone favors dynamical gap
generation or not.

\begin{figure}
\includegraphics[width=2.8in]{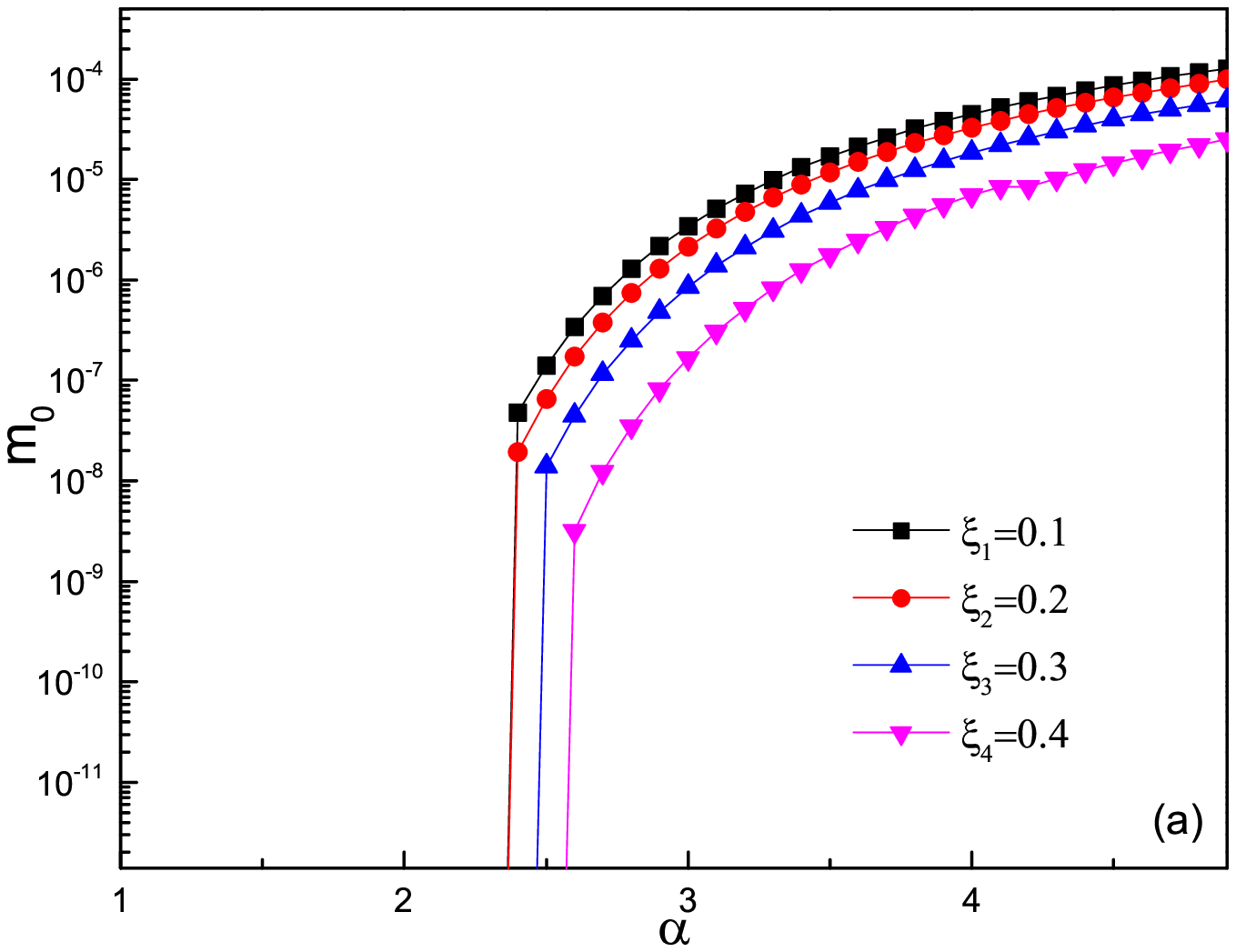}
\includegraphics[width=2.8in]{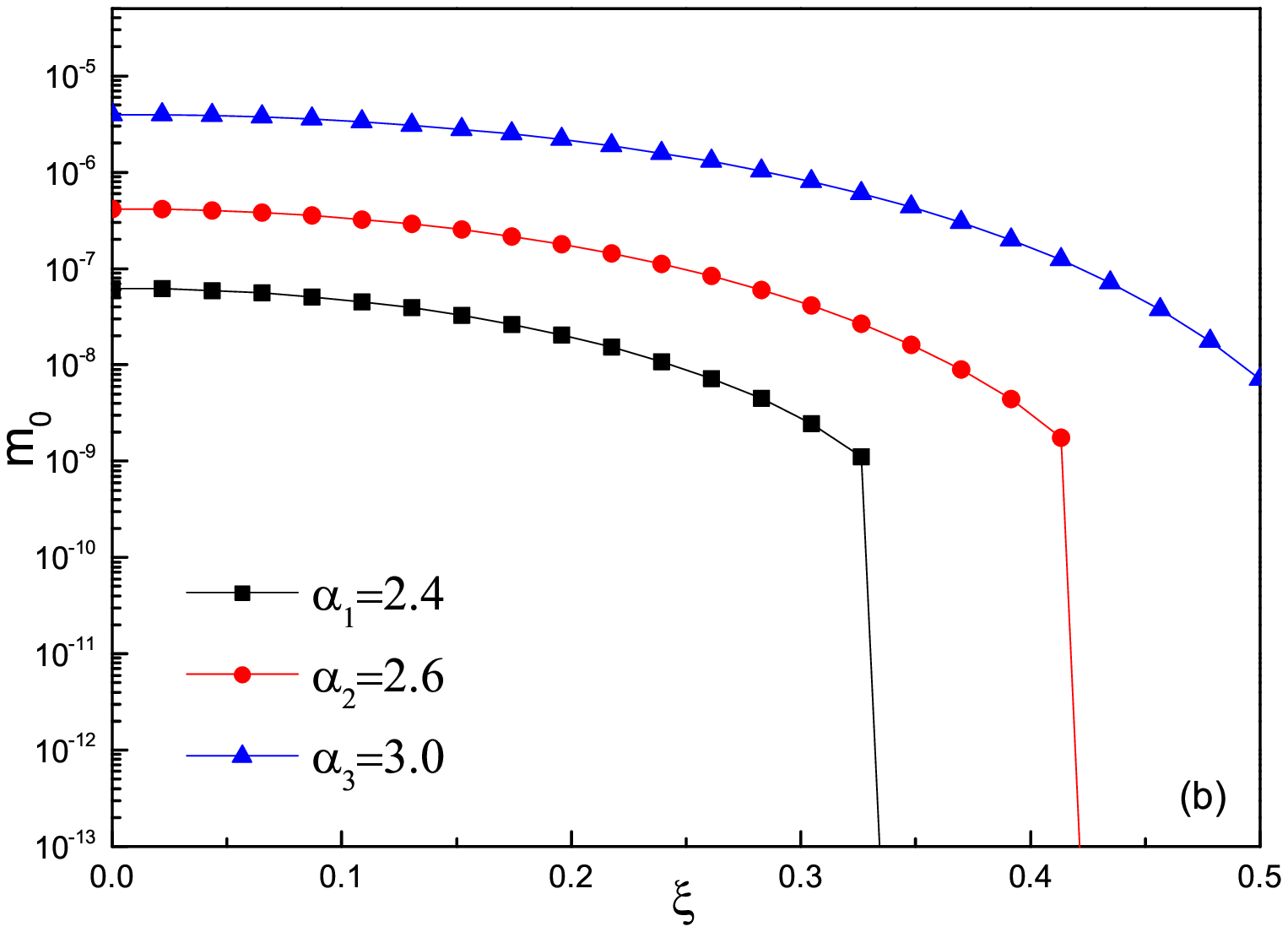}
\caption{(a) $\alpha$ dependence of $m_{0}$ for different $\xi$. (b)
$\xi$ dependence of $m_{0}$ for different $\alpha$.
\label{Fig:Gap0Tilted}}
\end{figure}

\begin{figure}
\includegraphics[width=2.8in]{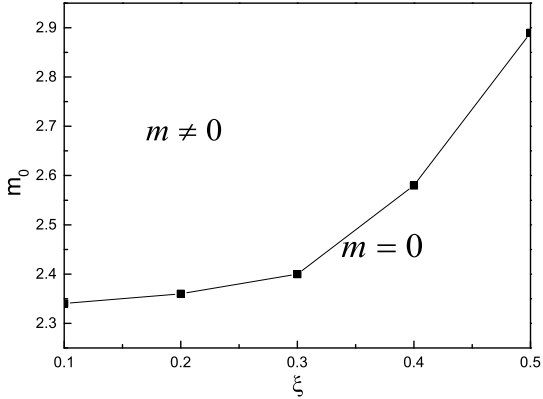}
\caption{Phase diagram of $\alpha$ and $\xi$ under instantaneous
approximation.}\label{Fig:PhaseDiagramTilted}
\end{figure}

The propagator of fermion excitations around a tiled Dirac cone has
the form \cite{Isobe12, Nishine10, Sari14, Trescher15, Proskurin15,
Hirata17}
\begin{eqnarray}
G(i\omega,\mathbf{k}) = \frac{1}{-i\omega\gamma_{0} +
v_{0}k_{x}\gamma_{0} + v\mathbf{k}\cdot\mathbf{\gamma}}.
\end{eqnarray}
We only consider the case $v_{0} < v$, so that the Fermi surface is
still discrete points. For simplicity, we assume that $v_{x} = v_{y}
= v$ and focus on the influence of tilt of  Dirac cone. The
polarization is defined as
\begin{eqnarray}
\Pi(i\Omega,\mathbf{q}) &=& -N_{f}\int\frac{d\omega}{2\pi}
\frac{d^2\mathbf{k}}{(2\pi)^{2}}
\mathrm{Tr}\big[\gamma_{0}G_{0}(i\omega,\mathbf{k})\gamma_{0}
\nonumber \\
&& \times G_{0}(i\omega+i\Omega,\mathbf{k}+\mathbf{q})\big].
\end{eqnarray}
According to Ref.~\cite{Nishine10}, under instantaneous
approximation, the polarization is given by
\begin{eqnarray}
\Pi(0,\mathbf{q}) &=& \frac{N_{f}}{8}\frac{|\mathbf{q}|}{v\sqrt{1 -
\left(\frac{v_{0}}{v}\right)^{2}\cos^2\theta_{\mathbf{q}}}}
\nonumber \\
&=& \frac{N_{f}}{8}\frac{|\mathbf{q}|}{v\sqrt{1 -
\left(\frac{v_{0}}{v}\right)^{2}\frac{q_{x}^{2}}{q_{x}^{2} +
q_{y}^{2}}}},
\end{eqnarray}
here $N_{f}=2$.
It is then easy to get a dressed Coulomb function
\begin{eqnarray}
V(\mathbf{q}) &=& \frac{1}{\frac{\kappa|\mathbf{q}|}{2\pi
e^{2}}+\Pi(\mathbf{q})}\nonumber \\
&=&\frac{1}{\frac{|\mathbf{q}|}{2\pi \alpha v} + \frac{N_{f}}{8}
\frac{|\mathbf{q}|}{v\sqrt{1-\left(\frac{v_{0}}{v}\right)^{2}
\frac{q_{x}^{2}}{q_{x}^{2}+q_{y}^{2}}}}},
\end{eqnarray}
where
\begin{eqnarray}
\alpha=\frac{e^2}{v\kappa}.
\end{eqnarray}
To the leading order, the gap equation is
\begin{eqnarray}
m(p_{x},p_{y})&=& \int\frac{d\omega}{2\pi} \int\frac{dk_{x}}{2\pi}
\int\frac{dk_{y}}{2\pi} \nonumber \\
&& \times \frac{m(k_{y},k_{y})}{\left(\omega + iv_{0}k_{x}\right)^2
+ v^{2}k_{x}^2 + v^{2}k_{y}^{2} + m^{2}(k_{x},k_{y})} \nonumber \\
&& \times \frac{1}{\frac{|\mathbf{q}|}{2\pi \alpha v} +
\frac{N_{f}}{8}\frac{|\mathbf{q}|}{v\sqrt{1 -
\left(\frac{v_{0}}{v}\right)^{2}
\frac{q_{x}^{2}}{q_{x}^{2}+q_{y}^{2}}}}},
\end{eqnarray}
where
\begin{eqnarray}
q_{x}=p_{x}-k_{x},\qquad  q_{y}=p_{y}-k_{y}.
\end{eqnarray}
Performing the integration of $\omega$ by using the contour integral
and residue theorem, we obtain
\begin{eqnarray}
m(p_{x},p_{y}) &=& \frac{1}{2} \int\frac{dk_{x}}{2\pi}
\int\frac{dk_{y}}{2\pi} \nonumber \\
&& \times \frac{m(k_{y},k_{y})}{\sqrt{v^{2}k_{x}^2 + v^{2}k_{y}^{2}
+ m^{2}(k_{x},k_{y})}}\nonumber \\
&& \times \frac{1}{\frac{|\mathbf{q}|}{2\pi \alpha v}
+\frac{N_{f}}{8}\frac{|\mathbf{q}|}{v\sqrt{1-\xi^{2}
\frac{q_{x}^{2}}{q_{x}^{2}+q_{y}^{2}}}}}, \label{Eq:GapEquationTilt}
\end{eqnarray}
with
\begin{eqnarray}
\xi = \frac{v_{0}}{v}.
\end{eqnarray}
Here, we use $\xi$ to measure to what extend the Dirac cone is
tilted. $\xi=0$ represents the perfect Dirac cone of graphene, and
the increase of $\xi$ stands for the increase of tilted degree of
Dirac cone.

The numerical solutions of Eq.~(\ref{Eq:GapEquationTilt}) are
depicted in Fig.~\ref{Fig:Gap0Tilted}, which clearly informs us that
$m_{0}$ decreases with growing $\xi$. Therefore, the tilt of Dirac
cone reduces the possibility of dynamical gap generation. The phase
diagram in $\alpha$-$\xi$ plane is presented in
Fig.~\ref{Fig:PhaseDiagramTilted}.

\section{Summary and discussion}\label{Sec:Summary}

In this paper, we have studied dynamical excitonic gap generation in
2D Dirac semimetal with deformed Dirac cone caused by anisotropy and
tilt of Dirac cone. Firstly, we studied dynamical gap generation
with anisotropy, which is likely caused by uniaxial strain or
periodic potentials. After solving the DS gap equation under three
different approximations, we find that the decrease of fermion
velocities supports dynamical gap generation, but the velocity
anisotropy tends to suppress dynamical gap generation. Subsequently,
we have considered the organic material
$\alpha$-(BEDT-TTF)$_{2}$I$_{3}$, which is also a 2D Dirac
semimetal, and found that dynamical gap generation is suppressed by
the tilt of the Dirac cone.  This shows
that the shape and geometry of Dirac cone is closely related to the
formation of excitonic mass gap generation, which might help to
explore concrete materials that can potentially exhibit excitonic
insulating transition.

In an uniaxially strained graphene, the fermion dispersion becomes
anisotropic. However, it is necessary to emphasize that uniaxial
strain leads to not only velocity anisotropy, but also an
enhancement of effective Coulomb interaction strength due to the
decrease of the mean value of fermion velocity. Dynamical gap generation is suppressed
by the former effect, but promoted by the latter one. Therefore, the
ultimate influence of uniaxial strain on dynamical gap generation
can only be determined by considering these two competitive effects
simultaneously. The fate of dynamical gap generation depends on the
actual values of $\alpha$ and $\eta$. By adopting the uniaxial
strain dependence of fermion velocities in Ref.~\cite{Choi10}, we
show that the overall effects of uniaxial strain can induce an
excitonic gap in suspended graphene once the strength of uniaxial strain is
over a certain value (namely $7.34\%$). This is in accordance with
the results of Sharma \emph{et al.} \cite{Sharma17}. While they have
a misinterpretation of the results. It is the decreasing the fermion
velocity dominates over the suppression effect caused by the growing
velocity anisotropy, and thus induces an excitonic gap, not the
velocity anisotropy contributes to the formation of excitonic gap
generation.

Our analysis suggest that, a more efficient way to realize excitonic
insulating transition is to merely increase the interaction strength
by reducing the fermion velocity, without introducing any
anisotropy. For instance, one could apply a uniform and isotropic
strain on graphene, as suggested in Ref.~\cite{Tang15}.

\acknowledgments

We would thank Guo-Zhu Liu for valuable suggestions. This work is
supported by the National Natural Science Foundation of China under
Grants Nos. 11475085, 11535005, 11690030, and 11504379, the Natural
Science Foundation of Jiangsu Province under Grant No. BK20130387,
and the Jiangsu Planned Projects for Postdoctoral Research Funds
under Grant No. 1501035B.

\end{document}